# Local Density of States and Interface Effects in Semimetallic ErAs Nanoparticles Embedded in GaAs

Jason K. Kawasaki,[1] Rainer Timm,[2] Kris T. Delaney,[3] Edvin Lundgren,[2] Anders Mikkelsen,[2] and Chris J. Palmstrøm[1,4]

[1] Materials Department, University of California, Santa Barbara, CA 93106
[2] Nanometer Structure Consortium (nmC@LU), Department of Physics, Lund University, Sweden
[3] Materials Research Laboratory, University of California, Santa Barbara, California 93106, USA
[4] Department of Electrical and Computer Engineering, University of California, Santa Barbara, California 93106, USA

## ABSTRACT

The atomic and electronic structures of ErAs nanoparticles embedded within a GaAs matrix are examined via cross-sectional scanning tunneling microscopy and spectroscopy (XSTM/XSTS). The local density of states (LDOS) exhibits a finite minimum at the Fermi level demonstrating that the nanoparticles remain semimetallic despite the predictions of previous models of quantum confinement in ErAs. We also use XSTS to measure changes in the LDOS across the ErAs/GaAs interface and propose that the interface atomic structure results in electronic states that prevent the opening of a band gap.

The electronic properties of low dimensional semimetals and semiconductors are of great importance for a wide range of topics, from applications in nanostructured thermoelectric materials [1] to fundamental studies of topological insulators [2]. In these materials, quantum size effects are expected to produce significant changes from the bulk electronic band structure. For example, bulk HgTe is a semimetal with a band overlap of 150 meV [3]. But when its dimensions are confined to near the 2D limit, HgTe quantum become topological insulators, characterized by a topological $Z_2$ invariant [2,4]. When confined even further, extremely thin HgTe 2D quantum wells [4] and 0D nanoparticles [3] undergo a quantum confinement-induced semimetal-to-semiconductor transition.

ErAs is another technologically important semimetal, as it has been shown to grow epitaxially on III-As semiconductors with the As-sublattice remaining continuous across the interface [5,6]. Bulk ErAs has rocksalt crystal structure and is a semimetal with valence band maximum at $\Gamma$ and conduction band minimum at X. However given its relatively large $\Gamma$-X band overlap of $\Delta$ = 700 meV [7] (compare to 150 meV for HgTe [3,4]), the role of quantum confinement in determining its electronic band structure is much less certain. Indeed, for ultrathin ErAs films embedded in GaAs, simple effective mass models predicted that quantum confinement would open a band gap for films of thickness 1.73 nm (6 monolayers, ML) or less [8]. However, magnetotransport [8] and angle resolved photoemission spectroscopy measurements [9] have shown that such films remain semimetallic for thicknesses as low as 0.86 nm (3 ML).

For ErAs nanoparticles the confinement effects are expected to be even stronger, resulting from the reduced dimensionality of the nearly 0D nanoparticles. ErAs nanoparticles embedded within GaAs exhibit optical absorption peaks in the near infrared region [10], and one interpretation is that the absorption results from transitions across a confinement-induced band gap [7]. Based on this interpretation, Scarpulla *et al.* proposed a simple hard-walled finite-potential model that predicted a gap opening for embedded ErAs nanoparticles with diameters of approximately 3 nm [7]. However, given the failures of effective mass model for ErAs thin films [8,9], this hard-walled finite potential model has remained controversial. An alternative explanation is that the nanoparticles remain semimetallic, with the absorption resulting from excitation of surface plasmon resonances [10]. A direct measurement of the electronic structure of embedded ErAs nanoparticles is still needed in order to determine the validity of the models.

In this Letter, we report the first direct measurements of the electronic structure of ErAs nanoparticles embedded within a semiconducting GaAs matrix. We employ cross-sectional scanning tunneling microscopy (XSTM) and spectroscopy (XSTS). The embedded ErAs nanoparticle samples were grown by molecular beam epitaxy using (001) n-type GaAs substrates. The sample structure consists of four layers of varying coverage of ErAs (0.125, 0.25, 0.5, and 1.0 ML) separated by 125 nm n-type GaAs spacers. All layers were grown at a substrate temperature of 540°C with a constant Si doping of roughly 5 x $10^{18}$ cm$^{-3}$. Further growth details are described elsewhere [11].

After growth, the samples were cleaved in ultrahigh vacuum to expose a clean {110} surface [11] and analyzed at room temperature in an Omicron variable temperature scanning tunneling microscope. XSTS was performed by interrupting the feedback and simultaneously measuring the tunneling current (I) and the differential conductance (dI/dV) as a function of voltage (V) at specified points on the {110} surface. The conductance was measured using a lock-in amplifier with a 30 mV, 1.3 kHz modulation on the tip-sample bias. In order to amplify the conductance signal and gain a greater dynamic range, spectroscopy measurements were performed in variable gap mode [11,12]. To remove the tip-sample distance dependence, dI/dV was normalized by the absolute conductance I/V, which we have broadened by convolution with an exponential function in order to avoid divergence at the band gap [12]. After normalization, the quantity $(dI/dV)/(\overline{I/V})$ is proportional to the local density of states (LDOS), where the sample voltage corresponds to energy, in eV, referenced to the Fermi level [12].

Figures 1(a)-1(c) show representative filled states XSTM images of the ErAs nanoparticles in the low coverage limit of less than 0.5 ML. The vertical lines are As atomic rows on the GaAs {110} surface. Since the {110} is not the rocksalt ErAs cleavage plane, the particles tend not to cleave [11]. Instead, the particles remain stuck in one of the cleavage surfaces and are pulled out of the other. This results in protruding particles [Fig. 1(a)] or holes due to missing particles [Fig. 1(b)] in the cross-sectional STM images. The corresponding height profiles are shown in Fig. 1(d).

A histogram of particle lengths for the protruding and pulled-out particles is shown in Fig. 1(e). The particles appear nearly spherical, with average lengths of roughly 2.4 and 2.3 nm along the ⟨110⟩ and [001] directions, respectively. The 2.4 nm length along ⟨110⟩ is consistent with Kadow *et al.* [13], who measure a 2 nm diameter in the (001) plane for particles grown at a similar temperature. Thus the particles are clearly within the sub-3nm regime where hard-walled potential models predict a band gap [7].

A buried ErAs nanoparticle is shown in Fig. 1(c), with the corresponding height profile in Fig. 1(d). Here we see a smooth profile 0.07 nm in height overlaid on the atomic corrugation. This profile is Gaussian in shape with a standard deviation of $\sigma = 4.1$ nm and full width at half maximum of 4.8 nm. The apparent height further reduces from 0.07 nm to 0.05 nm when the sample bias is changed from -1.8 to -2.0 V. The small apparent height (less than one atomic step) and strong bias dependence suggest that this profile results from an electronic rather than a topographical feature. It is interpreted to be a buried ErAs particle whose electronic states induce electronic changes in the surrounding GaAs matrix, such as band bending or introduction of localized states into the GaAs band gap.

XSTS measurements were performed in order to further explore the electronic structure of the embedded ErAs nanoparticles. Figure 2(a) shows normalized dI/dV spectra for the GaAs matrix and protruding ErAs nanoparticles. Both curves are averaged over at least 20 individual spectra. In the GaAs spectra a clear band gap extending from -1 to 0.8 V is observed. Because of tip-induced band bending the measured band gap of 1.8 eV is larger than the true band gap of 1.4 eV, consistent with previous STS studies [14,15]. Additionally, despite the heavy n-type doping (5 x $10^{18}$ cm$^{-3}$ Si) the GaAs Fermi level is pinned near midgap, which is often observed for metal-GaAs interfaces [16,17] and for cleaved surfaces due to atomic steps [18].

The ErAs nanoparticle dI/dV shows no evidence of a band gap. Instead, dI/dV (LDOS) exhibits a sharp but finite minimum at the Fermi level, indicating that the nanoparticles are semimetallic. This curve is qualitatively similar to density functional theory (DFT) calculations for the bulk ErAs density of states [19]. Additionally, spectra measured directly over buried particles [Fig. 1(c)] are nearly identical to spectra measured over protruding particles [Fig. 1(a)]. Thus the observed semimetallic behavior is not induced by cleavage defects or the vacuum interface, but is instead a feature of the particles themselves. These measurements suggest that the observed near-IR optical absorption is probably not due to optically driven electron-hole excitations, but instead results from the excitation of surface plasmons. This lies in direct contrast with the simple hard-walled potential model, which predicts that 2.3 nm spherical particles should have a band gap on the order of 0.5 eV [7].

The local electronic features across the interface between ErAs and GaAs may also influence this behavior. Fig. 2(b) shows a series of individual normalized dI/dV spectra starting at a point directly on top of a nanoparticle and moving in steps of 1.3 nm along the ⟨110⟩ direction into the GaAs matrix. Directly on top of the ErAs particle (0 nm) and near the particle edge (1.3 nm) the spectra retain the finite minimum at the Fermi level, consistent with semimetallic behavior. In both curves there is clear evidence of an extra state, not derived from bulk GaAs or ErAs, at 0.2 eV, indicated by an arrow. Moving across the ErAs/GaAs interface to a distance of 2.6 nm, which is roughly 1.4 nm into the GaAs matrix, the state at 0.2 eV begins to decay and the minimum at the Fermi level broadens; however there are still states within the GaAs band gap close to the particle. These states continue to decay and the bulk GaAs DOS is recovered near a distance of 3.9 nm from the particle center. This 3.9 nm decay radius is in good agreement with the $\sigma = 4.1$ nm radius of electronic contrast for the buried particle observed by XSTM [Fig. 1(c)].

These states within the band gap, and, in particular, the state at 0.2 eV that decays with distance into the GaAs matrix, may result from interface states. Note that the state at 0.2 eV does not appear in DFT calculations for bulk ErAs [19] or in photoemission spectra of continuous ErAs films [20]. But for ErAs/GaAs interfaces, DFT calculations predict the existence of interface states for both (001) [17,21] and (110) planar interfaces [21] at positions within the GaAs band gap. These states arise from differences in bonding and coordination across the ErAs (rocksalt) / GaAs (zincblende) interface, and they peak at the interface and decay into the GaAs matrix, just as observed in our XSTS measurements. Here the decay occurs primarily into the GaAs side because in the case of a semimetal/semiconductor interface, the interface states correspond to extended states from the semimetal ErAs side [21].

These interface states may be responsible for preventing the opening of a band gap. For ErAs thin film superlattices on (001) GaAs, DFT calculations by Said *et al.* show that ErAs/GaAs interface states persist even with reduced ErAs film thickness, and their positions at and near the Fermi level prevent a gap from opening [22]. Additionally, tight binding calculations for GdAs/GaAs superlattices by Xia *et al.* [23] identify a heavy hole interface band along the Γ-X dispersion that curves up and turns into a conduction band. This partially filled interface band prevents GdAs/GaAs superlattices from turning into a semiconductor, and Xia *et al.* argue that the same may be true for ErAs/GaAs planar superlattices.

Similar mechanisms may prevent ErAs nanoparticles from opening a band gap; however for the case of embedded nanoparticles, the interfaces are more complicated than the simple (001) and (110) planar interfaces.

A potential effect of the observed interface states is to effectively reduce the size of the confining potential over some length scale into the GaAs matrix. Following Scarpulla et al. [7], we begin modeling the confinement using a spherically symmetric step potential whose height is given by the energy differences in the band extrema for GaAs and ErAs (Fig. 3 inset). The potential height for holes is $U_{0,h} = \Gamma_{VB,ErAs} - \Gamma_{VB,GaAs} = 1.03$ eV and for electrons is $U_{0,e} = X_{CB,GaAs} - X_{CB,ErAs} = 1.47$ eV. Note we used the room temperature band gap for GaAs, whereas Scarpulla et al. used the 0 K band gap. Our effective masses were $m^*_h/m_0 = 0.5$ (0.235) and $m^*_{e,X}/m_0 = 0.32$ (0.25) for GaAs (ErAs) [7].

We next apply two modifications to the finite-step potential model to include (1) the effects of interface states and (2) many-body effects (Fig. 3 insert). In the first modification we model an interface state as an intermediate step in the confinement potential with energy $E_{int}$ and spatial extent $d_{int}$. From XSTS measurements this state is located at approximately $E_{int} = 0.2$ eV above the Fermi level, and from DFT [22] and XSTS we find that the state is highly localized at the interface with width on the order of $d_{int} = a_{GaAs}$ (lattice constant of GaAs, 5.65 Å). The resulting interface step potential has the form $U_{step}(r) = 0$ for $r < a$, $U_{step}(r) = U_{0,int}$ for $a < r < a + d_{int}$, and $U_{step}(r) = U_{0,e/h}$ for $r > a + d_{int}$, where $a$ is the radius of the spherical ErAs nanoparticle.

To model many-body effects at the interface we note that the semimetallic nature of bulk ErAs, and the potential presence of surface plasmons at the ErAs/GaAs interface, motivate a Thomas-Fermi-like screening of the confining potential of the form $U_{screen}(r) = 1$ for $r < a$ and $U_{screen}(r) = -\exp[-k_{eff}(r-a)]+1$ for $r > a$, where $k_{eff}$ is the effective screening wave number. For an electron density of $5 \times 10^{18}$ cm$^{-3}$ the Thomas-Fermi wave number is 3.57 nm$^{-1}$, and we use this to guide the order of magnitude of the screening wave vector: $k_{eff} = 1$ nm$^{-1}$. The total confinement potential is given by $U_{total,e/h}(r) = U_{step,e/h}(r)U_{screen}(r)$, where we have adjusted $U_{0,int}$ such that after multiplying by the screening, $U_{total,e/h}(a+d_{int}) = E_{int}$ (Fig. 3 insert).

We next solve the Schrödinger equation in spherical coordinates to find the band shifts of occupied electron and hole states subject to this confining potential. The confinement-induced ErAs band gap is given by $E_g(a) = E_e(a) + E_h(a) - \Delta$. The results for the modified model with interface states and screening are shown in Fig. 3.

We find that compared to the simple hard-walled step-potential model, the presence of features associated to interface states and metallic screening provides a strong modification to the predicted confinement-induced gap opening. With these effects, at 2.3 nm diameter the particles are predicted to remain semimetallic, consistent with our XSTS measurements. Furthermore, when solved for a 2D thin film, the interface and screening model predicts that ErAs films should remain semimetallic down to a critical thickness of 0.15 nm. This 0.15 nm thickness is much less than the 1 ML (0.287 nm) physical limit, indicating that ErAs thin films will in fact never become semiconducting, consistent with previous experimental [8,9] and DFT [22,23] work on ErAs thin films.

Our analysis clearly demonstrates that the choice of the form of the confining potential has a strong effect on the predictions of simple one-electron confinement models. Our results also highlight the importance of including physically motivated features of the interface electronic structure in modeling the subtle effects of quantum confinement, especially in systems where differences in bonding and crystal structure across the interface lead to highly localized interface states. However, we caution that the results of such simple models are strongly dependent on the choice of parameters. For example, a choice of $k_{eff} = 0.5$ nm$^{-1}$ instead of 1 nm$^{-1}$ with the same values of $E_{int}$ and $d_{int}$ yields a band gap opening at 1.5 nm diameter instead of roughly 2.2 nm. Thus while these modifications may capture more of the complex interfacial physics, they also motivate future theoretical work of fully atomistic and parameter-free calculations to provide a truly quantitative understanding of the effects of quantum confinement in ErAs/GaAs.

In conclusion, we have examined the atomic and electronic structures of ErAs nanoparticles embedded within GaAs (001) via XSTM/XSTS. Tunneling spectroscopy shows that the LDOS of the ErAs particles has a sharp but finite minimum at the Fermi level, demonstrating that the particles are semimetallic. The data strongly suggest that previously observed optical absorption is due to surface plasmon resonances and that the simple hard-walled potential model does not provide an accurate description of quantum confinement for embedded ErAs nanoparticles. Tunneling spectroscopy also shows a state at 0.2 eV above the Fermi level that decays with distance across the ErAs/GaAs interface, and we attribute this to an interface state. We have shown that small changes to the model potential, motivated by the presence of interface states and metallic screening, strongly modify the predictions of the model and provide agreement with measurements, demonstrating the importance of considering the atomistic and electronic structure of the interface itself.

We thank A. C. Gossard, C. G. Van de Walle, and J. Buschbeck for helpful discussions. This work was supported by NSF through the IMI Program (No. DMR 0843934) and the UCSB MRL (No. DMR 05-20415), ARO (Award No. W911NF-07-1-0547), AFOSR (No. FA9550-10-1-0119), Swedish Research Council (VR), Swedish Foundation for Strategic Research (SSF), Crafoord Foundation, Knut and Alice Wallenberg Foundation, and European Research Council under European Union's Seventh Framework Programme (FP7/2007-2013)-ERC Grant agreement No. 259141. J. K. K. acknowledges funding from NDSEG. R. T. acknowledges support from the European Commission under a Marie Curie Intra-European Fellowship.

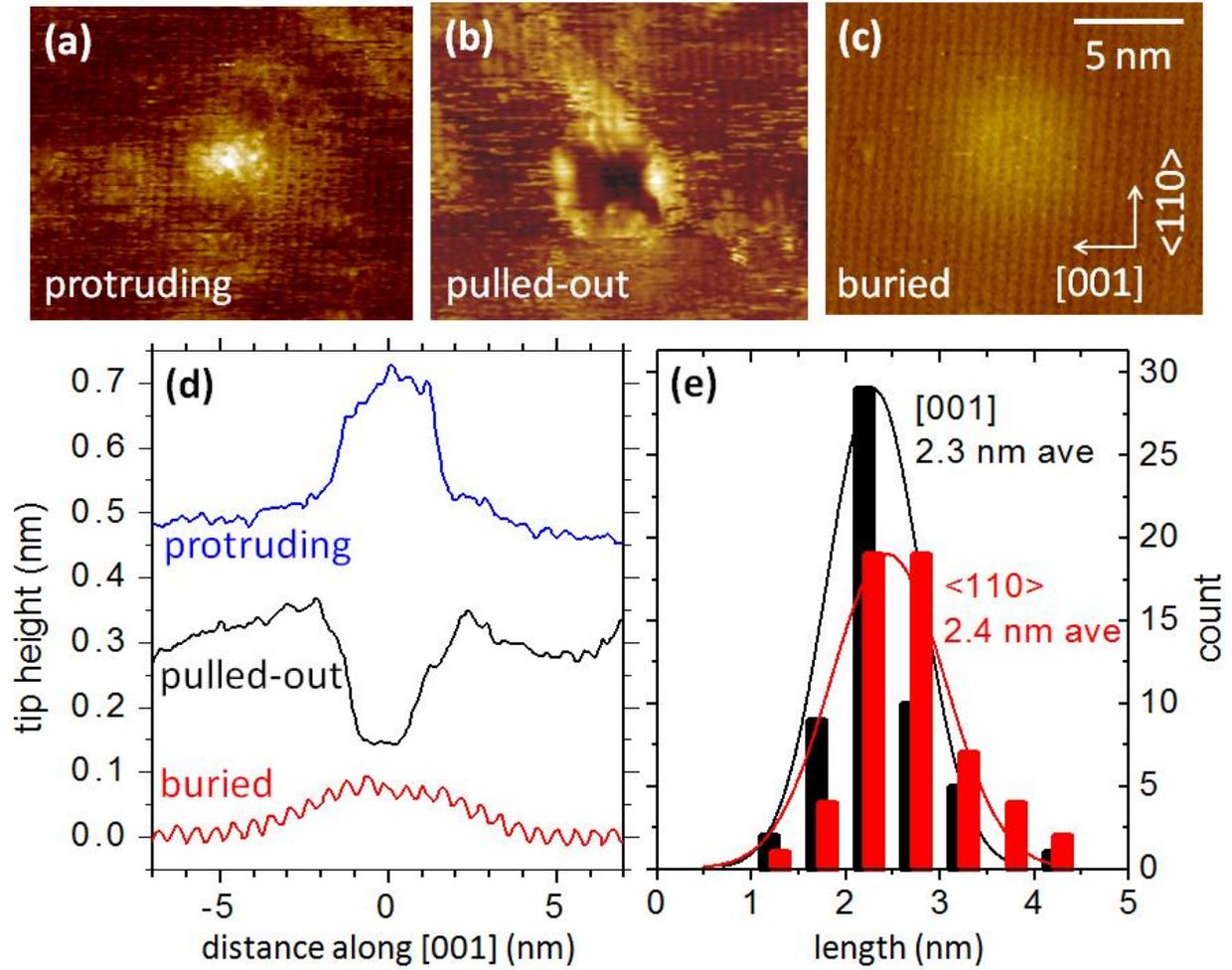

Figure 1. (color online) Filled states XSTM images of (a) protruding, (b) pulled-out, and (c) buried ErAs nanoparticles grown on (001) GaAs. (d) Height profiles of the three particle sites. (e) Histogram of the protruding and pulled-out particle lengths along [001] and ⟨110⟩.

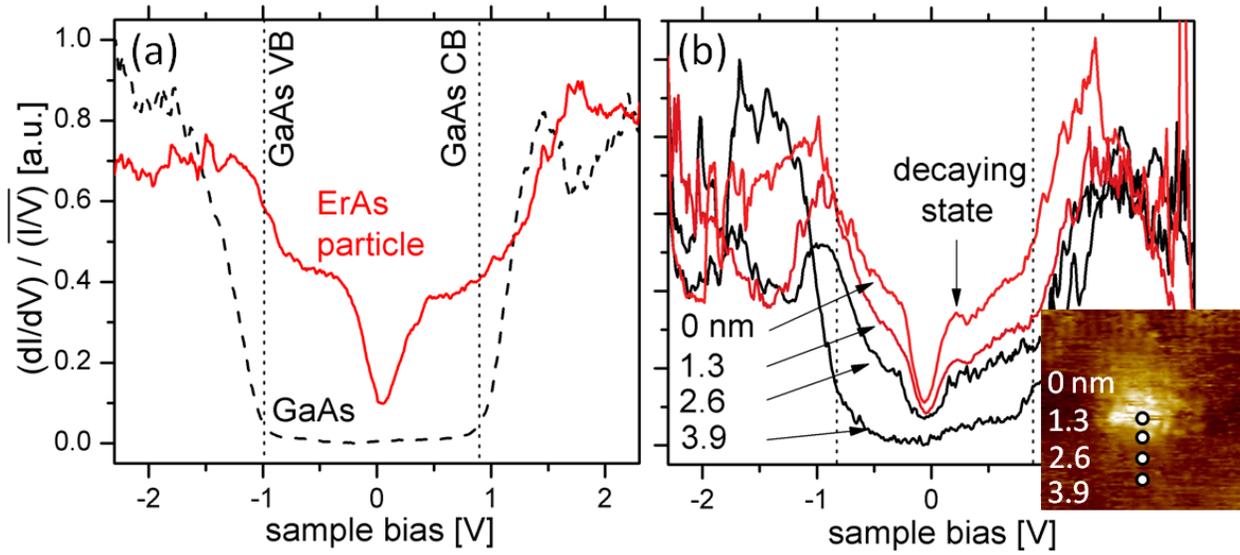

Figure 2. (color online) (a) Averaged differential conductance curves for protruding ErAs nanoparticles and the GaAs matrix. (b) Individual differential conductance spectra at varying points directly on top of an ErAs particle (0 nm) and moving in steps of 1.3 nm into the GaAs matrix (3.9 nm).

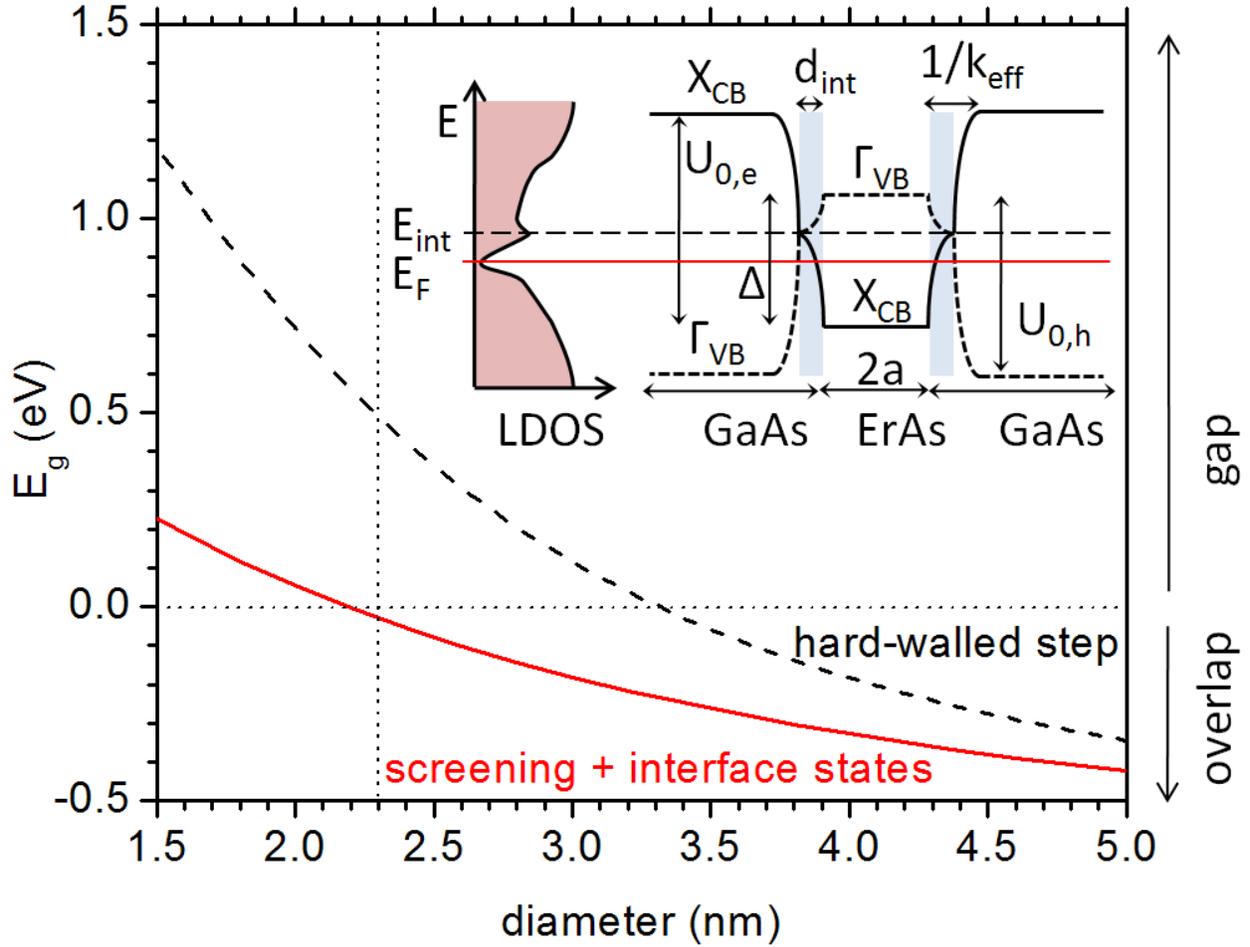

Figure 3. (color online) Calculated energy gap versus ErAs particle diameter. Insert shows schematic of the modified confinement potential model.